\documentclass[a4paper,10pt]{article}
\usepackage{ae}
\usepackage[ansinew]{inputenc}
\usepackage{hyperref}
\usepackage{lscape}
\usepackage{graphicx}
\usepackage{verbatim}
\usepackage{amssymb,amsmath,amsthm}

\newcounter{mnotecount}[section]

\newcommand{\mnotex}[1]
{\protect{\stepcounter{mnotecount}}$^{\mbox{\footnotesize $\bullet$\themnotecount}}$ 
\marginpar{
\raggedright\tiny\em
$\!\!\!\!\!\!\,\bullet$\themnotecount: #1} }

\newtheorem{thm}{Theorem}

\newtheorem{remark}{Remark}

\theoremstyle{plain}

\newtheorem{definition}{Definition}
\newtheorem{proposition}{Proposition}

\theoremstyle{remark}

\title{On the conservation of Superenergy and its applications}
\author{Alfonso Garc\'{\i}a-Parrado G\'omez-Lobo\thanks{E-mail: alfonso@math.uminho.pt}\\\\
{\small Centro de Matem\'atica, Universidade do Minho, 4710-057 Braga,
Portugal}}

\date{\today}

\begin{document}

\maketitle

\begin{abstract}
In this work we present a geometric identity involving the Bel-Robinson tensor which is formally similar to the Sparling
identity (which involves the Einstein tensor through the Einstein 3-form). In our identity the Bel-Robinson tensor enters
through the {\em Bel-Robinson 3-form} which, we believe, is introduced in the literature for the first time.
The meaning of this identity is that it is possible to formulate a {\em generic} conservation law for the quantity represented
by the Bel-Robinson tensor (superenergy). We also show how one can use the Bel-Robinson 3-form to estimate the components
of the Bel-Robinson tensor which are computed with respect to the causal elements of a frame. This estimate could be useful in a global existence
proof of the solutions of a theory of gravitation in dimension four.
\end{abstract}


\section{Introduction}
Conservation laws in Physics play a very important role in the sense that they enable us to draw important conclusions 
about the dynamics of a physical system (and sometimes even to determine the dynamics itself completely). There is 
an interesting connection between conservation laws and the invariance under symmetry transformations established 
by the Noether theorem. Roughly speaking this result states that if the action of a mechanical system is invariant
under a symmetry group, then one can construct a number of {\em conserved currents} out of the generators of the 
group. If one attempts to apply this result to the case of General Relativity, it turns out that the action is 
invariant under the group of diffeomorphisms and since the set of generators of the group of 
smooth diffeomorphisms is the set of smooth vector fields we conclude that it must be possible to construct
a conserved current out of any smooth vector field. In this case the physical quantity being conserved is the 
{\em energy} and therefore one 
can state that energy is always conserved in General Relativity. It is possible to distinguish two contributions 
to the {\em total energy}: the part arising from the matter energy-momentum tensor and the part coming from the 
{\em gravitational energy}. The former is represented by a tensor (the energy-momentum tensor) whereas 
the latter is only pseudo-tensorial. For this reason it is said that gravitational energy is pseudo-tensorial; 
it cannot be represented by a covariant energy momentum tensor in the same way as the matter energy-momentum can. 

Conservation laws do also play a role in Mathematics, even though their relation to a physical dynamics might not 
be apparent (or necessary). For example in a proof of uniqueness a conservation law usually plays a key role. Therefore the construction
of a conservation law is always important from a mathematical point of view. As it is well-known the mathematical 
expression of a conservation law is a vector field whose divergence vanishes (conserved current) 
because then the conservation law can be expressed in integral form through the Gauss theorem.

One of the aims of this paper is to show how one can construct {\em generic} conserved currents involving the Bel-Robinson 
tensor. This is a tensor which has important positivity properties in Lorentzian geometry and it has already played a role in the proof 
of important results in General Relativity (see e.g. \cite{CHRKLA93,CHRISTODOULOU-COLLAPSE,BONILLA-SENOVILLA}). 
The point is that in order to obtain conserved currents out of the Bel-Robinson tensor contracted with a set of vector fields one
needs to add other additional terms in a fashion similar to what one does when constructing conserved currents out of the energy-momentum tensor.
We show that some of the additional terms are covariant and some other are not. This follows 
a pattern similar to the conservation of energy in General Relativity. Since the physical quantity supposedly represented
by the Bel-Robinson tensor has been called {\em superenergy} in the literature we refer to our construction as the {\em conservation
of superenergy}. The physical status of superenergy has not reached a commonly accepted interpretation 
and while we do not come up with a new proposal in this regard, we show that one can formulate a conservation law 
for superenergy in General Relativity which is mathematically similar to the conservation of energy. To the best of our knowledge
this is a new result.
   
Another new result presented in this work is given in theorem \ref{thm:estimate} and it is summarized as follows: 
if we can find a foliation of our region of interest which is consistent with a condition (\ref{eq:estimate}) 
defined by a frame in a certain way, 
then the components of the Bel-Robinson tensor with respect to the causal frame elements can be estimated when integrated on the leaves of the foliation
in terms of the components integrated on a leave playing the role of the ``initial data hypersurface''. 
In fact we can use the frame itself to construct the foliation if we take one of the frame elements as the normal vector to the leaves 
so we can regard the consistency condition as a condition imposed on a frame (gauge choice).

The conservation of superenergy has been already the subject of some research. In 
\cite{SUPERENERGY} it was established the conservation in the {\em exchange} of superenergy between 
the Einstein-Klein-Gordon field and the gravitation and at a characteristic hypersurface 
of the electromagnetic field in a gravitational background). An interesting question is the possibility 
of superenergy exchange between the {\em gravitational superenergy}, represented by the Bel-Robinson tensor, and the 
{\em electromagnetic superenergy}, represented by the Chevreton tensor \cite{CHEVRETON}. To test this possibility 
one needs to construct conserved quantities involving the Bel-Robinson tensor and the Chevreton tensor. 
This possibility has been explored in \cite{ERIKSSON-CURRENT-II,ERIKSSON-CURRENT-I,ERIK-BERG-SENO,LAZKOZ-VERA-SENO} 
assuming the existence of Killing vectors with certain geometric properties.

The outline of this paper is as follows: we review the conservation of energy in General Relativity 
in section \ref{sec:sparling}. In a modern mathematical language this conservation law is encoded in 
the {\em Sparling identity}.  In section \ref{sec:bel-robinson} we introduce the {\em Bel-Robinson 3-form} 
which appears to be a new mathematical object. The Bel-Robinson 3-form enables us to write the Bel-Robinson tensor 
in an appropriate way. The most important parts of this work are section 
\ref{sec:superenergy-conservation} 
where we present a geometric identity (eq. (\ref{eq:superenergy-conservation}))
involving the Bel-Robinson 3-form which bears a formal resemblance to the Sparling identity and 
and section \ref{sec:estimate} where the estimate of the components of the Bel-Robinson tensor is derived.

Most of the computations of this paper have been done with the system {\em xAct} \cite{XACT}.

\section{The Sparling identity and the conservation of energy}
\label{sec:sparling}
Let $(M,{\boldsymbol g})$ be a 4-dimensional space-time (signature convention $(-,+,+,+)$). Latin small letters shall 
be used to denote numerical indices ranging from 1 to 4. 
The manifold $M$ defines the tangent bundle $T(M)$ and the bundle of frames $L(M)$ in the standard way. A point $u$ of $L(M)$ 
can be written as the pair $u=(x,\{{\boldsymbol u}_i(x)\})$ where $x=\pi(u)$, $\pi: L(M)\rightarrow M$ is the bundle projection and 
$\{{\boldsymbol u}_i(x)\}$ is a basis of $T_x(M)$. Let $\{\hat{\boldsymbol u}^j(x)\}$ be the dual basis of $\{{\boldsymbol u}_i(x)\}$.
Then any tensor field defined on $M$ induces a set of scalar functions on $L(M)$ obtained by taking its components in the bases
$\{{\boldsymbol u}_i(x)\}$, $\{\hat{\boldsymbol u}^j(x)\}$. For example the metric $\boldsymbol{g}$ defines the functions 
$g_{ab}(u)={\boldsymbol g}({\boldsymbol u}_a(x),{\boldsymbol u}_b(x))$ if 
$u=(x,\{{\boldsymbol u}_i(x)\})\in L(M)$. Similarly, 
the inverse metric ${\boldsymbol g}^\#$ and the volume element ${\boldsymbol \eta}$
induce the functions 
$g^{ab}(u)={\boldsymbol g}^\#(\hat{\boldsymbol u}^a(x),\hat{\boldsymbol u}^b(x))$ and 
$\eta_{abcd}(u)=\sqrt{-g(u)}\epsilon_{abcd}$, where $g(u)=\mbox{det}(g_{ab}(u))$ and 
$\epsilon_{abcd}$ is the alternating symbol with $\epsilon_{1234}=+1$.
Unless otherwise stated, tensor fields on $M$ will be regarded as sets of scalar quantities on $L(M)$ defined in this fashion.

Let us introduce now a set of 1-form 
fields $\{{\boldsymbol\theta}^a\}$ (canonical forms) in $L(M)$ defined by 
${\boldsymbol\theta}^a|_u(\vec{\boldsymbol X})=A^a$ 
$\Leftrightarrow$ $\pi_*(u)(\vec{\boldsymbol X})=A^a {\boldsymbol u}_a(x)$, 
$\forall\vec{\boldsymbol X}\in T_u(L(M))$. The Hodge dual acts 
as follows on a basis of forms constructed with elements of the set $\{{\boldsymbol\theta}^a\}$ 
\begin{eqnarray}
&&*({\boldsymbol\theta}^a)=\frac{1}{3!}\eta^{a}_{\phantom{a}bcd}{\boldsymbol\theta}^b\wedge {\boldsymbol\theta}^c\wedge {\boldsymbol\theta}^d
\;,\quad *({\boldsymbol\theta}^a\wedge{\boldsymbol\theta}^b)=
\frac{1}{2}\eta^{ab}_{\phantom{ab}cd}{\boldsymbol\theta}^c\wedge {\boldsymbol\theta}^d\;,\nonumber\\
&&*({\boldsymbol\theta}^a\wedge{\boldsymbol\theta}^b\wedge{\boldsymbol\theta}^c)=\eta^{abc}_{\phantom{abc}d}{\boldsymbol\theta}^d\;,
\quad *({\boldsymbol\theta}^a\wedge{\boldsymbol\theta}^b\wedge{\boldsymbol\theta}^c\wedge{\boldsymbol\theta}^d)=
\eta^{abcd}\;,
\label{eq:hodge}
\end{eqnarray}
where here and in other places we use $g^{ab}$ (resp. $g_{ab}$) 
to raise (lower) indices of indexed quantities. 

A torsion-free connection on $L(M)$ is determined by a set of 1-forms $\Gamma^a_{\phantom{a}b}$ on $L(M)$
fulfilling the {\em Cartan structure equations}
\begin{equation}
d({\boldsymbol\theta}^a)=-\Gamma^a_{\phantom{a}b}\wedge{\boldsymbol\theta}^b\;,\quad
d(\Gamma^a_{\phantom{a}b})=\mathcal{R}_{b}^{\phantom{b}a}-\Gamma^a_{\phantom{a}c}\wedge\Gamma^c_{\phantom{c}b}\;,\quad
d(\mathcal{R}_b^{\phantom{b}a})=\mathcal{R}_{c}^{\phantom{c}a}\wedge\Gamma^{c}_{\phantom{c}b}-\Gamma^{a}_{\phantom{a}c}
\wedge\mathcal{R}_{b}^{\phantom{b}c}\;,
\label{eq:cartan}
\end{equation}
The set $\Gamma^a_{\phantom{a}b}$ determines the connection 1-forms 
and the set $\mathcal{R}_{c}^{\phantom{c}a}$ the curvature 2-forms.
We have that
\begin{equation}
 \mathcal{R}_a^{\phantom{a}b}=\frac{1}{2} R_{cda}^{\phantom{cda}b}{\boldsymbol\theta}^c\wedge{\boldsymbol\theta}^d\;,
\label{eq:riemann-form}
\end{equation}
where $R_{cda}^{\phantom{cda}b}$ is the Riemann tensor and $R_{ca}\equiv R_{cda}^{\phantom{cda}d}$ is the Ricci tensor.
In addition if the connection is the Levi-Civita connection, one has the extra compatibility condition
\begin{equation}
 d (g_{ab})=g_{ac}\Gamma^c_{\phantom{c}b}+g_{cb}\Gamma^c_{\phantom{c}a}\;,
\label{eq:levi-civita}
\end{equation}
where $g_{ab}$ is regarded as a set of 0-forms on $L(M)$ (scalar fields on $L(M)$). 
From this, we deduce
\begin{equation}
 d(\eta_{abcd})=\eta_{fbcd}\Gamma^f_{\phantom{f}a}+\eta_{afcd}\Gamma^f_{\phantom{f}b}+
\eta_{abfd}\Gamma^f_{\phantom{f}c}+\eta_{abcf}\Gamma^f_{\phantom{f}d}.
\label{eq:diff-eta}
\end{equation}

The integrability conditions of (\ref{eq:cartan})-(\ref{eq:levi-civita}) are
\begin{equation}
0=-\mathcal{R}^{ab}\wedge{\boldsymbol\theta}_b\;,\quad
0=\mathcal{R}_{ab}+\mathcal{R}_{ba}\;,
\label{eq:cartan-integ-conditions} 
\end{equation}

Let us now introduce the following vector subspaces of $T_u(L(M))$
\begin{equation}
H_u\equiv\{\vec{\boldsymbol X}\in T_u(L(M)):\Gamma^a_{\phantom{a}b}|_{u}(\vec{\boldsymbol X})=0\}\;,
V_u\equiv\{\vec{\boldsymbol X}\in T_u(L(M)):{\boldsymbol\theta}^a|_{u}(\vec{\boldsymbol X})=0\}\;,
\end{equation}
$H_u$ is called the {\em horizontal subspace} and $V_u$ is the {\em vertical subspace}. It is clear that these form distributions 
on $L(M)$, denoted respectively by $H$ and $V$. Also one has that $T_u(L(M))=H_u\oplus V_u$. Note that $V_u$ can be defined generically 
on $L(M)$ whereas the definition of $H_u$ requires the introduction of the connection $\Gamma^a_{\phantom{a}b}$.
Consider now the following definition taken from \cite{KOBAYASHI}
\begin{definition}
Let $\tilde{R}_g:L(M)\rightarrow L(M)$ be the standard right action of $GL(4,\mathbb{R})$ on $L(M)$. For $m\in\mathbb{N}$,
a $\mathbb{R}^m$-valued differential form ${\boldsymbol\omega}$ on $L(M)$ of degree $r$ is called pseudo-tensorial 
if $\tilde{R}^*_g{\boldsymbol\omega}=\rho(g^{-1}){\boldsymbol\omega}$, where $g\in GL(4,\mathbb{R})$ and 
$\rho:GL(4,\mathbb{R})\rightarrow\mathbb{R}^m$ is a representation of the group $GL(4,\mathbb{R})$ on $\mathbb{R}^m$.
The form ${\boldsymbol\omega}$ on $L(M)$ is called tensorial (or horizontal) if in addition to the previous condition one has
that ${\boldsymbol\omega}(\vec{\boldsymbol X}_1,\cdots,\vec{\boldsymbol X}_r)=0$ whenever at least one of the vectors 
$\vec{\boldsymbol X}_i$ is vertical. 
\end{definition}

The sets of differential forms on $L(M)$ introduced so far can be regarded as pseudo-tensorial when 
regarded as forms with values in $\mathbb{R}^m$ for appropriate $m$. 
In addition the forms, ${\boldsymbol\theta}^a$, 
$\mathcal{R}_{a}^{\phantom{a}b}$ are examples of tensorial forms. The main difference 
between pseudo-tensorial and tensorial forms is that
the pullback of a pseudo-tensorial form by a local 
section $\sigma:M\rightarrow L(M)$ induces a form on $M$ which does not transform
covariantly whereas this does not happen with a tensorial form. From the physical point of view tensorial 
forms are expected to be related to 
{\em covariant quantities} whereas pseudo-tensorial forms are related to frame-dependent quantities.
A $R^{m}$-valued differential form on $L(M)$ will be just called a form.
Sometimes, tensorial forms shall be referred to simply as tensors.

With this geometric set-up, we can define the Nester-Witten 2-form
\cite{FRAUENDIENER-PSEUDO,SZABADOS-PSEUDO,DUBOIS-VIOLETTE}.
\begin{equation}
 L_{a}\equiv \frac{1}{2}\eta_{abcd}{\boldsymbol\theta}^b\wedge\Gamma^{cd}\;,\quad
\end{equation}
Using (\ref{eq:cartan})-(\ref{eq:diff-eta}) one gets after a computation
that the 2-form $L_{a}$ fulfill the identity (Sparling identity)

\begin{equation}
d L_a=\mathcal{E}_a+\mathcal{S}_a\;,\quad
\mathcal{E}_a\equiv\frac{1}{2}\eta_{abcd}{\boldsymbol\theta}^b\wedge\mathcal{R}^{cd}\;,\quad
\mathcal{S}_a\equiv\frac{1}{2}(\eta_{bcde}{\boldsymbol\theta}^b\wedge\Gamma^c_{\phantom{c}a}-
\eta_{abce}{\boldsymbol\theta}^b\wedge\Gamma^{c}_{\phantom{c}d})\wedge\Gamma^{de}
\label{eq:sparling}
\end{equation}
See \cite{FRAUENDIENER-PSEUDO,DUBOIS-VIOLETTE,SZABADOS-PSEUDO} for the details of this computation. 
From the Sparling identity we deduce  
\begin{equation}
\delta (*\mathcal{E}_a+ *\mathcal{S}_a)=0\;, 
\label{eq:energy-conservation}
\end{equation}
where $*$ is the Hodge dual and $\delta$ the co-differential. Recall that if ${\boldsymbol\omega}$ is a 1-form in $L(M)$ then
the pull-back of $\delta{\boldsymbol\omega}$ to $M$ by a local section $\sigma:M\rightarrow L(M)$ yields
\begin{equation}
\sigma^*(\delta{\boldsymbol\omega})=\mbox{div}(\sigma^*{\boldsymbol\omega})\;, 
\end{equation}
where ``div'' represents the divergence operator computed with respect to the metric ${\boldsymbol g}$ (now 
regarded as a tensor field on $M$). 
In this sense eq. (\ref{eq:energy-conservation})
can be understood as a conservation law which tells us that the sum of $*\mathcal{E}_a$ and $*\mathcal{S}_a$ 
always gives
a conserved current when pulled back to $M$ by a local section. The 1-form $*\mathcal{E}_a$ (Einstein 1-form) 
is a tensorial 1-form given by
\begin{equation}
*\mathcal{E}_a=-G_{ab}{\boldsymbol\theta}^b\;,\quad G_{ab}\equiv R_{ab}-\frac{1}{2}g_{ab} R.
\end{equation}
Hence, the pull-back of $*\mathcal{E}_a$ to $M$ can be regarded as the energy-momentum flux of the matter 
if we use the Einstein's equations.
The 1-form $*\mathcal{S}_a$ is pseudo-tensorial but not tensorial and thus, its pull-back to 
$M$ will depend on the section chosen to define 
the pull-back. The current obtained after performing the pull-back can be interpreted as the flux of gravitational 
energy. Therefore the physical 
content of  eq. (\ref{eq:energy-conservation}) is that the energy-momentum flux of the matter plus the 
gravitational energy flux is always conserved 
(energy conservation). The energy-momentum flux is covariant as it is the pull-back of 
a tensorial form whereas the gravitational energy
flux is not because it is the pull-back of a pseudo-tensorial form. This is the standard interpretation 
that one cannot define a covariant 
``energy momentum tensor'' for the gravitational field. The pull-back of 
$\mathcal{S}_a$ for a number of specific choices of a section has been 
computed in the literature \cite{FRAUENDIENER-PSEUDO,SZABADOS-PSEUDO}, 
showing that classical pseudo-tensors used to represent 
the energy-momentum of the gravitational field are recovered in this way
 
From (\ref{eq:sparling}) 
and (\ref{eq:cartan})-(\ref{eq:diff-eta})
one gets after a computation
\begin{equation}
d(\mathcal{S}_a)=-\mathcal{E}_d\wedge\Gamma^d_{\phantom{d}a}\;,\quad
d(\mathcal{E}_a)=\mathcal{E}_{b}\wedge\Gamma^b_{\phantom{b}a}.
\label{eq:integrability-sparling}
\end{equation}
These equations can be regarded as the {\em integrability conditions} of the Sparling identity.

\section{The Bel-Robinson 3-form}
\label{sec:bel-robinson}
The Weyl 2-form $\mathcal{W}_a^{\phantom{a}b}$ is defined as follows
\begin{equation}
\mathcal{W}_a^{\phantom{a}b}\equiv\mathcal{R}_a^{\phantom{a}b}+{\boldsymbol\theta}^b\wedge S_a-{\boldsymbol\theta}_a\wedge S^b\;,\quad
S_a\equiv {\boldsymbol\theta}^b S_{ba}.
\label{eq:weyl-form}
\end{equation}
Here $S_{ab}$ is the Schouten tensor
\begin{equation}
S_{ab}\equiv\frac{1}{2}\left(R_{ab}-\frac{1}{6}g_{ab} R\right)\;,\quad R\equiv g^{ab}R_{ab} 
\label{eq:schouten}
\end{equation}
From (\ref{eq:weyl-form}) one deduces
\begin{equation}
\mathcal{W}_a^{\phantom{a}b}=\frac{1}{2} W_{cda}^{\phantom{cda}b}{\boldsymbol\theta}^c\wedge{\boldsymbol\theta}^d\;,\quad
\mathcal{W}_{ab}+\mathcal{W}_{ba}=0\;, 
\label{eq:weyl-properties}
\end{equation}
where $W_{cda}^{\phantom{cda}b}$ is the Weyl tensor. Define now the following 1-form 
\begin{equation}
\Omega^{de}_{\phantom{de}abc}\equiv \frac{1}{2}{\rm g}_{bc}*(\mathcal{W}^{de}\wedge\theta_a)+
2{\delta}^{e}_{\phantom{e}(b}*(\mathcal{W}_{c)}^{\phantom{c}d}\wedge\theta_a)
\label{eq:omega}
\end{equation}
If we define the Weyl tensor right dual (which equals the Weyl tensor left dual in dimension 4) by
\begin{equation}
W^*_{\phantom{*}abcd} \equiv \tfrac{1}{2} \eta_{cd}{}^{eh} W_{abeh}\;,
\end{equation}
then one finds that (\ref{eq:omega}) can be rendered in the form
\begin{equation}
\Omega^{de}_{\phantom{de}abc}
=\boldsymbol{\theta}^{h} (- 2\delta_{(c}{}^{e} (W^*)^{d}{}_{b)ah}+ 
\tfrac{1}{2}g_{bc} (W^*)^{de}{}_{ah})\;,
\label{eq:omega-tensor}
\end{equation}
where (\ref{eq:hodge}) and (\ref{eq:weyl-properties}) have to be used.
The Bel-Robinson 3-form is defined by 
\begin{equation}
 \mathcal{T}_{abc}\equiv \Omega_{\phantom{ed}abc}^{ed}\wedge\mathcal{W}_{ed}.
\label{eq:define-br-form}
\end{equation}
\begin{proposition}
One has the relation
\begin{equation}
*(\mathcal{T}_{abc})=\theta^d T_{abcd}\;, 
\label{eq:br-dual}
\end{equation}
where $T_{abcd}$ is the Bel-Robinson tensor
\begin{equation}
T_{ablm} \equiv W_{ajmk} W_{b}{}^{j}{}_{l}{}^{k} + W_{ajlk} W_{b}{}^{j}{}_{m}{}^{k} -  
\tfrac{1}{8} g_{ab} g_{lm} W_{dfgh}W^{dfgh}.
\end{equation}
\end{proposition}
\proof An explicit computation using (\ref{eq:weyl-properties}) and (\ref{eq:hodge}) shows that
\begin{eqnarray}
&& *(*(\mathcal{W}_{da}\wedge{\boldsymbol\theta}_b)\wedge\mathcal{W}^d_{\phantom{d}c})=-{\boldsymbol\theta}^d W_{afde}
W^{\phantom{b}e\phantom{c}f}_{b\phantom{e}c}+\frac{1}{8}{\boldsymbol\theta}_b g_{ac} W_{defh}W^{defh}\;,\label{eq:weyl-weyl1}\\
&&*(*(\mathcal{W}_{ba}\wedge{\boldsymbol\theta}_d)\wedge\mathcal{W}^d_{\phantom{d}c})=
\frac{1}{2}{\boldsymbol\theta}_b g_{ac} W_{defh}W^{defh}\label{eq:weyl-weyl2}\;,
\end{eqnarray}
where the dimensionally dependent identity
\begin{equation}
W_{ahcd}W_b^{\phantom{b}hcd}=\frac{1}{4}g_{ab}W_{ihcd}W^{ihcd}\;, 
\label{eq:weyl-ddi}
\end{equation}
was used along the way. Combining eqs. (\ref{eq:define-br-form}) and (\ref{eq:omega}) one gets
\begin{equation}
 *(\mathcal{T}_{abc})=-*(*(\mathcal{W}_{db}\wedge{\boldsymbol\theta}_a)\wedge\mathcal{W}^d_{\phantom{d}c})-
*(*(\mathcal{W}_{dc}\wedge{\boldsymbol\theta}_a)\wedge\mathcal{W}^d_{\phantom{d}b})+
\frac{1}{2}g_{bc}*(*(\mathcal{W}_{e}^{\phantom{e}d}\wedge{\boldsymbol\theta}_a)\wedge\mathcal{W}^e_{\phantom{e}d})\;,
\label{eq:BR-dual}
\end{equation}
which upon using (\ref{eq:weyl-weyl1})-(\ref{eq:weyl-weyl2}) leads to (\ref{eq:br-dual}).\qed

If we recall that the Bel-Robinson tensor is totally symmetric and traceless then the previous proposition leads to
\begin{equation}
\mathcal{T}_{(abc)}=\mathcal{T}_{abc}\;,\quad \mathcal{T}^a_{\phantom{a}ac}=0. 
\end{equation}

We present another mathematical property of the Bel-Robinson 3-form.
\begin{proposition}
\begin{equation}
\mathcal{T}_{abc}\wedge{\boldsymbol\theta}^a=0.
\label{eq:BRwedgetheta}
\end{equation}
\label{prop:BRwedgetheta}
\end{proposition}
\proof To prove this we need to take into account the identity
\begin{equation}
*(\mathcal{W}_{da}\wedge{\boldsymbol\theta}_{b})\wedge\mathcal{W}_{fc}\wedge{\boldsymbol{\theta}}^{h}=
(-W_{ad}{}^{he} W_{becf} + \tfrac{1}{2} \delta_{b}{}^{h} W_{ad}{}^{eg} W_{cfeg}){\boldsymbol\theta}^1\wedge{\boldsymbol\theta}^2
\wedge{\boldsymbol\theta}^3\wedge{\boldsymbol\theta}^4\;,
\end{equation}
 which is obtained by an explicit computation. One can now expand the left hand side of (\ref{eq:BRwedgetheta}) using the Hodge dual of 
(\ref{eq:BR-dual}) and work out the resulting terms with the previous identity. The final result follows after using the cyclic 
property $W_{abcd}+W_{acdb}+W_{adbc}=0$ and the dimensionally dependent identity (\ref{eq:weyl-ddi}).\qed

Since its introduction in \cite{BEL-RADIATION-STATE} 
the Bel-Robinson tensor has been extensively studied in the literature (see \cite{SUPERENERGY} and references
therein for a thorough review). Specially important are the {\em positivity properties} of this tensor which 
state that for any set of 
causal, future-directed vector fields $\{\vec{\boldsymbol{u}}_1,\vec{\boldsymbol{u}}_2,
\vec{\boldsymbol{u}}_3,\vec{\boldsymbol{u}}_4\}$ on $M$ one has that the {\em super-energy density}
${\boldsymbol T}(\vec{\boldsymbol{u}}_1,\vec{\boldsymbol{u}}_2,\vec{\boldsymbol{u}}_3,\vec{\boldsymbol{u}}_4)$
is non-negative. Moreover, the super-energy density vanishes for a set of timelike vector fields if and only if 
the Weyl tensor is zero. These and other properties bear some resemblances with the mathematical properties the energy-momentum 
tensor of a physical system has and therefore the possible physical interpretation of the Bel-Robinson tensor has been
the subject of much debate during the years. In geometric units the physical dimensions of the Bel-Robinson tensor
are $L^{-4}$ with $L$ standing for length, and hence they do not correspond to energy. For that reason the 
purported physical quantity represented by the Bel-Robinson tensor has been termed as {\em superenergy}.
We show in the next section that one can formulate for the superenergy a {\em conservation law} similar 
to that found in eq. (\ref{eq:energy-conservation}) for the total energy (gravitational plus matter).

\section{The conservation of superenergy}
\label{sec:superenergy-conservation}
In this section we show that there exists an identity formally similar to (\ref{eq:sparling}) involving 
the Bel-Robinson tensor. To start with we define the {\em superenergy potential}
\begin{equation}
 \mathcal{Z}_{abc}\equiv\Omega^{de}_{\phantom{de}abc}\wedge\Gamma_{de}.
\label{eq:defineZ}
\end{equation}
Using the Cartan equations (\ref{eq:cartan}) and the properties of the exterior derivative, one easily finds
\begin{equation}
d \mathcal{Z}_{abc}=(d(\Omega^{de}_{\phantom{de}abc})+
\Gamma^{e}_{\phantom{e}f}\wedge\Omega^{df}_{\phantom{af}cde})\wedge\Gamma_{de}
+\Omega^{de}_{\phantom{ed}abc}\wedge\mathcal{R}_{ed}\;,
\end{equation}
Writing in the last equation the curvature $\mathcal{R}_{ed}$ in terms of the Weyl 2-form 
$\mathcal{W}_{ed}$ by means of (\ref{eq:weyl-form}) and using (\ref{eq:define-br-form}) we get
\begin{equation}
d \mathcal{Z}_{abc}=\Xi^{de}_{\phantom{ed}abc}\wedge\Gamma_{de}+\mathcal{K}_{abc}+\mathcal{T}_{abc}\;,
\label{eq:superenergy-conservation}
\end{equation}
where
\begin{equation}
\mathcal{K}_{abc}\equiv 2\Omega^{[ed]}_{\phantom{[ed]}abc}\wedge S_e\wedge\theta_d\;,\quad
\Xi^{ab}_{\phantom{ab}cde}\equiv d(\Omega^{ab}_{\phantom{ab}cde})
+\Gamma^{b}_{\phantom{b}f}\wedge\Omega^{af}_{\phantom{af}cde} 
\label{eq:define-Xi}
\end{equation}
From eq. (\ref{eq:superenergy-conservation}) we deduce
\begin{equation}
0=\delta *d \mathcal{Z}_{abc}=\delta\left(*(\Xi^{de}_{\phantom{de}abc}\wedge\Gamma_{de})+*(\mathcal{K}_{abc})+*(\mathcal{T}_{abc})\right).
\label{eq:superenergy-current-conservation}
\end{equation}
This equation is formally similar to (\ref{eq:energy-conservation}) and it expresses the fact that the combination of the currents 
arising from the pull-backs to $M$ of the 1-forms $*(\Xi^{de}_{\phantom{ab}abc}\wedge\Gamma_{de})$, $*(\mathcal{K}_{abc})$ and 
$*(\mathcal{T}_{abc})$ by any section is \underline{always} conserved. Note that $\mathcal{T}_{abc}$ and $\mathcal{K}_{abc}$ are tensorial forms 
whereas $\Xi^{de}_{\phantom{ab}abc}$ is pseudo-tensorial. This means that a general conservation law of the superenergy requires the presence 
of terms defined by means of pseudo-tensors, in a way similar to that of the conservation of energy.
In this sense eq. (\ref{eq:superenergy-conservation}) can be regarded as the counterpart of the Sparling's identity 
(\ref{eq:sparling}) when dealing with superenergy.

We have already explained the relevance of the 3-form $\mathcal{T}_{abc}$ and its relation to the Bel-Robinson tensor 
in section \ref{sec:bel-robinson}.
As for the 3-form $\mathcal{K}_{abc}$ a computation using (\ref{eq:weyl-form}) and (\ref{eq:omega}) shows that 
\begin{equation}
 *(\mathcal{K}_{abc})={\boldsymbol\theta}^j K_{jabc}\;,
\end{equation}
where $K_{jabc}$ is defined by 
\begin{eqnarray}
&& K_{jabc}\equiv
S_{c}{}^{d}W_{abjd} + S_{b}{}^{d} W_{acjd}+ S_{a}{}^{d}W_{bdcj} +S_{a}{}^{d} W_{bjcd}-  g_{bc} S^{de}W_{adje}- \nonumber\\
&& - 
2g_{aj}S^{de}W_{bdce} +g_{ac} S^{de} 
W_{bdje} + g_{ab} S^{de} W_{cdje}.
\end{eqnarray}
By construction $K_{ja(bc)}=K_{jabc}$.
Also an explicit computation shows that 
$K_{j\phantom{a}ac}^{\phantom{j}a}=0$, $K_{ja\phantom{b}b}^{\phantom{ja}b}=0$ which implies that 
$\mathcal{K}_{abc}$ is traceless. In addition 
$\mathcal{K}_{abc}$ can be decomposed in the way
\begin{equation}
\mathcal{K}_{abc}=\mathcal{N}_{cab}+\mathcal{N}_{bac}+\Sigma_{abc}\;,\quad \mathcal{N}_{cab}\equiv 
\frac{1}{3}(\mathcal{K}_{abc}-\mathcal{K}_{bac})\;,\quad
\Sigma_{abc}\equiv \mathcal{K}_{(abc)}.
\label{eq:decompose-K}
\end{equation}
Note also that 
\begin{equation}
\mathcal{N}_{[abc]}=0.
\end{equation}
This means that the 3-form $\mathcal{N}_{abc}$ can be regarded as a {\em Lanczos candidate} because it has 
the same algebraic properties of the Lanczos potential of the Weyl tensor.
The conservation law (\ref{eq:superenergy-conservation}) then entails
\begin{equation}
d \mathcal{Z}_{[ab]c}=\Xi^{de}_{\phantom{ed}[ab]c}\wedge\Gamma_{de}-\frac{3}{2}\mathcal{N}_{cab}\;,\quad
d \mathcal{Z}_{(abc)}=\Xi^{de}_{\phantom{ed}(abc)}\wedge\Gamma_{de}+\mathcal{T}_{abc}+\Sigma_{abc}.
\end{equation}
From this equation we conclude that one can construct a conservation law involving only $\mathcal{N}_{abc}$ as the 
tensor part of the conservation. 
Similarly, one sees that the object $\Sigma_{abc}$ also appears as the tensor part in a conservation law but 
it always does so in combination with the Bel-Robinson 3-form. Therefore $\Sigma_{abc}$ 
can be regarded as the responsible of the interaction between the matter and the superenergy of the gravitation.

For completeness we compute the integrability conditions of (\ref{eq:superenergy-conservation}).
Taking the exterior derivative of this expression and using Cartan equations (\ref{eq:cartan}) 
we get 
\begin{equation}
 d(\mathcal{T}_{abc})+d(\mathcal{K}_{abc})=(\Xi^{de}_{\phantom{de}abc}-
\Omega^{df}_{\phantom{df}abc}\wedge\Gamma^{e}_{\phantom{e}f})\wedge\mathcal{R}_{de}.
\end{equation}
If we use (\ref{eq:define-Xi}) this can be re-written in the form
\begin{equation}
d(\mathcal{T}_{abc}+\mathcal{K}_{abc})=
\Psi^{de}_{\phantom{de}abc}\wedge\mathcal{R}_{de}\;,\quad
\Psi^{de}_{\phantom{de}abc}\equiv
d\Omega^{de}_{\phantom{de}abc}-\Omega^{df}_{\phantom{de}abc}\wedge\Gamma^{e}_{\phantom{e}f}
-\Omega^{fe}_{\phantom{fe}abc} \wedge\Gamma^d_{\phantom{d}f}.
\label{eq:def-psi}
\end{equation}
This equation can be regarded as the counterpart of (\ref{eq:integrability-sparling}).

\section{A new estimate involving the Bel-Robinson superenergy}
\label{sec:estimate}
One of the most important applications of the Bel-Robinson tensor is in the proof of global existence results for certain inital data sets of the Einstein field 
equations (see \cite{CHRKLA93,CHRISTODOULOU-COLLAPSE,KLAINNICO03} for explicit applications). The idea in these proofs is to construct positive quantities
with integrals involving the components of the Bel-Robinson in carefully chosen frames and then use the quantities in estimates which yield the existence
results through a bootstrap argument. In constructing the positive quantities, the positivity properties of the Bel-Robinson tensor reveal themselves essential
(a thorough review and proof of these properties can be found in \cite{SUPERENERGY}). In a sense the Bel-Robinson tensor acts as a kind of 
norm for the gravitational field which vanishes if and only if the space-time is flat.

The choice of frames in which the estimates hold is in general a highly non-trivial task which has been carried out only in particular situations. 
In this sense no successful attempt has been made so far to find a general procedure which enables us to select a gauge in which the components of 
the Bel-Robinson tensor in that frame can be estimated although research in this direction has been certainly conducted \cite{MONCRIEFBR}.
In this section we prove a result (theorem \ref{thm:estimate}) which may certainly aid in this difficult task.

The Bel-Robinson 3-form has been obtained as a 3-form in the bundle of frames but in this section 
we shall choose a section $\sigma: M\rightarrow L(M)$ and use it to pull ${\mathcal T}_{abc}$ back to the base 
manifold. The section $\sigma$ will be assumed global and all the differential forms appearing in this section
will be understood as (tensor valued) forms on $M$ (elements of the de Rham complex $\Lambda(M)$). Since ${\mathcal T}_{abc}$
is a tensorial form, its pullback under $\sigma$ transforms covariantly and therefore we still use the symbol
${\mathcal T}_{abc}$ for the pull-backed 3-form. However, the indices $(abc)$ should now be regarded as component indices 
in the particular frame defined by the section $\sigma$.   

Our starting point is the following identity which is valid for any non-vanishing 
1-form ${\boldsymbol\omega}\in\Lambda^1(M)$ and any set of component indices $(abc)$ with 
respect to a given frame $\sigma$
\begin{equation}
 d{\mathcal T}_{abc}=\varphi\; {\boldsymbol\omega}\wedge {\mathcal T}_{abc}\;,\varphi\in C^{\infty}(M). 
\label{eq:main-identity}
 \end{equation}
This identity holds because its left and right hand sides are both 4-forms and $\Lambda^4(M)$ is a 1-dimensional 
module if $M$ is 4-dimensional.

\begin{proposition}
 If $(a,b,c)$ are frame indices corresponding to causal future-directed frame elements and $\Sigma\subset M$ is a 
spacelike or null smooth co-dimension 1 orientable submanifold (or a submanifold with points of both of these types) then 
\begin{equation}
 \int_{\Sigma}{\mathcal{T}_{abc}}\geq 0.
\label{eq:positive-integral}
 \end{equation}
\label{prop:positive}
\end{proposition}

\proof The integral of the 3-form $\mathcal{T}_{abc}$ can be transformed as follows

\begin{equation}
\int_{\Sigma}{\mathcal{T}_{abc}}=\int_{\Sigma}\langle *(\mathcal{T}_{abc}),\vec{\boldsymbol n}\rangle d\Sigma\;, 
\end{equation}
where $\vec{\boldsymbol n}$ is the unit normal to $\Sigma$ and $d\Sigma$ represents the volume form on $\Sigma$ defined
by 
$$
d\Sigma\equiv i_{\vec{\boldsymbol n}}{\boldsymbol\eta}\;,
$$
with ${\boldsymbol\eta}$ being the volume form of $M$. If $\Sigma$ is null, the normal $\vec{\boldsymbol n}$ has to be
chosen in such a way that $d\Sigma$ is a non-degenerate 3-form (the choice is not unique). Suppose now that the vector fields 
$\{\vec{\boldsymbol u}_a,\vec{\boldsymbol u}_b,\vec{\boldsymbol u}_c\}$ are the frame elements corresponding to 
the indices $(a,b,c)$. If we use (\ref{eq:br-dual}) to replace 
$*(\mathcal{T}_{abc})$ then the integral is rendered in the form
$$
\int_{\Sigma} {\boldsymbol T}(\vec{\boldsymbol u}_a,\vec{\boldsymbol u}_b,\vec{\boldsymbol u}_c,\vec{\boldsymbol n})d\Sigma\;,
$$
where ${\boldsymbol T}$ is the Bel-Robinson tensor. The hypotheses of the proposition imply that 
$\vec{\boldsymbol u}_a$, $\vec{\boldsymbol u}_b$, $\vec{\boldsymbol u}_c$ are all causal future-directed
vectors and we can always chose an orientation of $\Sigma$ which makes $\vec{\boldsymbol n}$ causal future directed too.
The proposition is now a direct consequence of the properties of the Bel-Robinson tensor (see considerations 
coming after the proof of proposition \ref{prop:BRwedgetheta}).
\qed

\begin{figure}[h]
\begin{center}
\includegraphics[width=.5\textwidth,keepaspectratio=true]{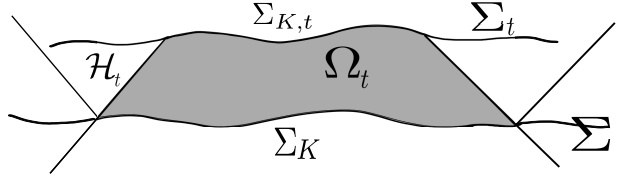}
\end{center}
\caption{\label{fig:figure} geometric construction used to define the set $\Omega_t$ (in dark grey).}
\end{figure}

\begin{thm}
Let $\Sigma\subset M$ be a 
spacelike smooth co-dimension 1 orientable submanifold and let 
$D^+(\Sigma)$ be its standard future Cauchy development. Assume that there exists a 
foliation of $D^+(\Sigma)$, $\{\Sigma_t\}_{t}$, $t\in I$, $I$ real interval  
and a frame $\sigma$ (section on $L(M)$) fulfilling the following properties
\begin{enumerate}
 \item $\Sigma_t$ is spacelike $\forall t\in I$ and $\Sigma\in\{\Sigma_t\}$.
 \item There are causal vector fields in the frame $\sigma$ (we shall take the label $(a,b,c)$ for the frame elements
 associated to these vector fields) such that 
 \begin{equation}
 0<\int_{\Sigma}{\mathcal T}_{abc}<\infty.
 \label{eq:initial-data}
 \end{equation}
 
 \item The scalar field $\varphi$ of the identity (\ref{eq:main-identity}) particularized for the frame indices 
 $(a,b,c)$ and the 1-form ${\boldsymbol\omega}=dt$ defined from the leaves of the foliation  has the property
\begin{equation}
\int_{\Sigma_t}\varphi\mathcal{T}_{abc}\leq m(t)\int_{\Sigma}\mathcal{T}_{abc}\;,\quad t\in I\;,
\label{eq:integrable-bound}
\end{equation}
where $m(t)$ is an integrable function in the interval $I$.
\end{enumerate}
Under all these assumptions the integral $\int_{\Sigma_t}{\mathcal T}_{abc}$ exists $\forall t\in I$.
\label{thm:estimate}
\end{thm}
\proof Without loss of generality, we can take  
$\Sigma=\Sigma_0$, $t>0$ in $\Sigma_t$. Let us consider a compact sub-manifold $\Sigma_{K,t}\subset\Sigma_t$
and define $\Sigma_K\equiv\Sigma\cap J^-(\Sigma_{K,t})$ (note that this is a non-empty subset as 
$\Sigma_{K,t}\subset D^+(\Sigma)$). We introduce now the 
set
\begin{equation}
 \Omega_t\equiv J^+(\Sigma_K)\cap J^-(\Sigma_{K,t}).
\end{equation}
The set $\Omega_t$ is compact \cite{HAWKING-ELLIS} and by construction we have 

\begin{equation}
\partial\Omega_t=\Sigma_K\cup\Sigma_{K,t}\cup\mathcal{H}_t\;, 
\label{eq:boundary-omega}
\end{equation}
where $\mathcal{H}_t\subset\partial J^-(\Sigma_{K,t})$.
The sub-manifolds $\Sigma_K$ and $\Sigma_{K,t}$ are space-like and 
$\mathcal{H}_t$ is null as it is a subset of $\partial J^-(\Sigma_{K,t})$ 
which is null \cite{SINGULARITY-REVIEW}.
See figure \ref{fig:figure} to 
get an intuition of this geometric construction. 

Relation (\ref{eq:boundary-omega}) and Stokes theorem entail
\begin{equation}
\int_{\Omega_t} d{\mathcal{T}_{abc}}=\int_{\partial{\Omega_t}}{\mathcal{T}_{abc}}=
\int_{\Sigma_{K,t}}\mathcal{T}_{abc}-\int_{\Sigma_K}\mathcal{T}_{abc}+
\int_{\mathcal{H}_t}\mathcal{T}_{abc}\geq \int_{{\Sigma}_{K,t}}\mathcal{T}_{abc}
-\int_{\Sigma_K}\mathcal{T}_{abc}\;,
\label{eq:int1}
\end{equation}
where we used in the last step proposition \ref{prop:positive} on the null $\mathcal{H}_t$ (note that 
that proposition implies that the other integrals are positive quantities too). Also, the 
identity (\ref{eq:main-identity}) applied to ${\boldsymbol\omega}=dt$ enables us to transform
the first integral
\begin{equation}
\int_{\Omega_t}d{\mathcal{T}_{abc}}=\int_{\Omega_t}\varphi\; dt\wedge\mathcal{T}_{abc}
=\int_0^{t}ds\int_{\Sigma_{K,s}}\varphi \mathcal{T}_{abc}.
\label{eq:int2}
\end{equation}
Note that all integrals in previous expressions are evaluated over compact sets and therefore
their existence is guaranteed. Combining eqs. (\ref{eq:int1})-(\ref{eq:int2}) we get
\begin{equation}
\int_{\Sigma_{K,t}}\mathcal{T}_{abc}\leq\int_{\Sigma_K}\mathcal{T}_{abc}+
\int_0^{t}ds\int_{\Sigma_{K,s}}\varphi\;\mathcal{T}_{abc}\leq
\left(1+\int_0^tm(s)ds\right)\int_{\Sigma_K}\mathcal{T}_{abc}\;,
\quad t\in I\;,
\label{eq:step-0}
\end{equation}
where in the last inequality we used (\ref{eq:integrable-bound}) and the integrability of $m(t)$. 
Finally, we note that this estimate holds for any compact  $\Sigma_{K,t}\subset\Sigma_t$
and that the quantity $m(t)$ is totally independent from the set $\Sigma_{K,t}$ we chose at 
the beginning. This fact and the assumption that $\int_{\Sigma}\mathcal{T}_{abc}$ exists
enables us to conclude that $\int_{\Sigma_t}\mathcal{T}_{abc}$ also exists $\forall t\in I$.
\qed
\begin{remark}\em
If we let $\Sigma_K$ approach $\Sigma$ then (\ref{eq:step-0}) becomes
\begin{equation}
 \int_{\Sigma_{t}}\mathcal{T}_{abc}\leq\left(1+\int_0^tm(s)ds\right)\int_{\Sigma}\mathcal{T}_{abc}\;,
\quad t\in I.
\label{eq:estimate}
\end{equation}
Using proposition \ref{prop:positive} we deduce that the term in brackets must be positive
\begin{equation}
0\leq 1+\int_0^tm(s)ds.
\end{equation}

\label{rem:estimate}
\end{remark}

\begin{remark}\em
 \label{rem:frame} We can chose a frame $\sigma$ such that it determines the foliation $\{\Sigma_t\}$ of theorem \ref{thm:estimate}
 (for example one of the elements of the frame is an integrable vector field which is normal to the leaves of the foliation). Under this 
 provision we can regard the conditions stated by theorem \ref{thm:estimate} as a set of conditions imposed on a frame (gauge choice). Therefore 
 theorem \ref{thm:estimate} is a geometric construction of a gauge with interesting properties. These properties may enable us to prove 
 global existence results for the Einstein equations as we explain below.
 
\end{remark}

 If the unit normal $\vec{\boldsymbol n}$ to the foliation $\{\Sigma_t\}$ is an element of the frame $\sigma$, then the estimate
 (\ref{eq:estimate}) can be written in the form
 \begin{equation}
  0<\int_{\Sigma_t}{\mathcal T}_{000}\leq \left(1+\int_0^tm(s)ds\right)\int_{\Sigma}{\mathcal T}_{000}\;,
 \end{equation}
where we chose the label 0 for the frame element associated to $\vec{\boldsymbol n}$. The computations
carried out in the proof of proposition \ref{prop:positive} enable us to re-write the previous equation as follows
\begin{equation}
 0<\int_{\Sigma_t}
 {\boldsymbol T}(\vec{\boldsymbol n},\vec{\boldsymbol n},\vec{\boldsymbol n},\vec{\boldsymbol n})
 d{\Sigma_t}\leq
\left(1+\int_0^tm(s)ds\right)\int_{\Sigma}
 {\boldsymbol T}(\vec{\boldsymbol n},\vec{\boldsymbol n},\vec{\boldsymbol n},\vec{\boldsymbol n})
 d\Sigma.
\label{eq:superenergy-estimate}
 \end{equation}
The scalar ${\boldsymbol T}(\vec{\boldsymbol n},\vec{\boldsymbol n},\vec{\boldsymbol n},\vec{\boldsymbol n})$
is the so-called superenergy density and from the considerations coming after the proof of proposition
\ref{prop:BRwedgetheta} we deduce that it is a strictly positive quantity which is zero if and only if the 
Weyl tensor is zero. Therefore we can regard the integrals appearing in equation 
(\ref{eq:superenergy-estimate}) as a measure of the strength of the gravitational field on $\Sigma$ and 
$\Sigma_t$. The mathematical properties of the superenergy density have been used successfully in
the proof of a number of mathematical results \cite{BONILLA-SENOVILLA,SE-DYNAMICALAWS,GAR-VAL-KERR,CHRKLA93,CHRISTODOULOU-COLLAPSE} and
we expect that the estimate shown in (\ref{eq:superenergy-estimate}) will be useful to prove global existence
results for the equations of any gravitational theory by means of a standard bootstrap argument (see e.g. \cite{TAODISPERSIVE}). 
If there exists a frame choice (gauge choice) $\sigma$ in which the assumptions of theorem \ref{thm:estimate} 
hold with $\mathcal{T}_{abc}$ as the superenergy density then 
we deduce from these results that any local existence result for any gravitational theory formulated in the gauge $\sigma$ could be turned into a global one. 
The idea is to construct inital data in which the condition (\ref{eq:initial-data}) is fulfilled because then (\ref{eq:superenergy-estimate}) implies that
the solution will be regular and bounded if $t\in I$ where $I$ represents the local existence interval. 
If it turns out that the function $m(t)$ is integrable on $\mathbb{R}$ (or an unbounded subset thereof) then it is plausible that
the solution can be extended by using repeatedly 
the local existence result and theorem \ref{thm:estimate} thus showing that the solution remains bounded and regular in the maximal data development.
In this sense it is important to formulate a local existence result in a gauge $\sigma$ and a foliation with the properties stated
by theorem \ref{thm:estimate}. This is the subject of ongoing work.

\section{Conclusions}
We have shown that one can construct conserved currents involving the classical Bel-Robinson tensor and other 
superenergy tensors in a manner which appears to be new. These conserved currents contain also a {\em pseudo tensorial part}
in a pretty much the same way as it happens when one considers conserved currents involving the energy-momentum 
tensor of the matter. As far as we are aware of, this is the first time that the pseudo-tensorial 
character of the superenergy is considered in the literature. We have shown that the conservation of superenergy 
adopts the form of a geometric identity which bears a formal similarity to the Sparling identity. 
In the latter case the energy of the matter enters in the identity via the Einstein equations whereas no 
counterpart to the Einstein equations exists in the conservation of the superenergy (there is no known dynamics for the superenergy). 
For this reason the conservation of the superenergy put forward in this paper should be regarded 
at this stage as a mathematical result rather than a physical law. 
An {\em independent} physical law of the 
{\em conservation of the superenergy} would exist if one could relate in a {\em physical fashion} the superenergy tensors 
$K_{abcd}$ and  $T_{abcd}$ to other superenergy tensors constructed from matter fields 
(for example the Chevreton tensor \cite{CHEVRETON} or similar). Of course we can always compute $K_{abcd}$ and 
$T_{abcd}$ for a solution of the Einstein equations and speak of the conservation of superenergy within the framework 
of the present work. However, in this particular case the conservation of superenergy is a derived law rather than 
an independent one.  We have also explored an independent application involving the construction of an estimate
of the {\em causal} components of the Bel-Robinson tensor with respect to a frame with certain properties. 
This estimate could be important in the proof of global existence results of arbitrary gravitational theories, not necessarily
based on the classical Einstein equations. 
Finally we should mention that our work has been restricted to dimension four. A possible 
generalisation to higher dimensions is under current research.

\section*{Acknowledgments}
We thank Prof. Jos\'e M. M. Senovilla for a careful reading of the manuscript and many valuable comments.
This work was supported by the Research Centre of Mathematics of the University of Minho (Portugal) 
with the Portuguese Funds from the ``Funda\c{c}\~ao para a Ci\^encia e a Tecnolog\'{\i}a (FCT)'', through the Project PEstOE/MAT/UI0013/2014
and through project CERN/FP/123609/2011.

\bibliographystyle{amsplain}
\bibliography{/home/alfonso/trabajos/BibDataBase/Bibliography}

\end{document}